\documentclass[10pt,a4paper]{article}
\usepackage{latexsym,epsfig,multicol}
\usepackage{color}
\usepackage{graphicx}
\usepackage{pifont}
\usepackage{verbatim}

\def\beq{\begin{eqnarray}}
\def\eq{\end{eqnarray}}

\def\lra{\longrightarrow}

\makeatletter

\newenvironment{figurehere}
  {\def\@captype{figure}}
  {}
\makeatother

\title{On signatures for the littlest Higgs model in electron-proton colliders}
\author{F.~M.~L.~de Almeida Jr., Y.~A.~Coutinho,, \\
J.~A.~Martins Sim\~oes, A.~J.~Ramalho and S.~Wulck.\\
Instituto de F\'isica\\
Universidade Federal do Rio de Janeiro\\
21945-970, RJ, Brazil\\
M.~A.~B.~do Vale\\
Depto. Ci\^encias Naturais\\
Universidade Federal de S\~ao Jo\~ao del Rei\\
36301-160, MG, Brazil}
\date{}

\begin{document}

\maketitle

\begin{abstract}
There is a recent proposal identifying the Higgs particle of the Standard
Model as a pseudo Nambu-Goldstone boson. This new broken symmetry introduces new
particles and new interactions. Among these new interactions a central role to get a
new physics is played by two new heavy neutral gauge bosons. We have studied
the new neutral currents in the Littlest Higgs model and compared with other extended models. For high energy $e^+ e^-$ colliders we present a clear signature for new neutral gauge bosons that can indicate the theoretical origin of these particles. Previous analysis by other authors were done at collider energies equal to the new gauge boson mass $M_{A_H}$. In this paper we show that asymmetries in fermion antifermion production can display model differences in the case
$M_{A_H} > \sqrt{s}$. For $M_{A_H} < \sqrt{s}$ we show that the hard photon energy distribution in $e^+ + e^- \lra \gamma + f + \bar f$ can present a model dependence. New bounds for the new neutral gauge boson masses are also presented.
\end{abstract}

\begin{multicols}{2}
The Higgs divergent radiative corrections indicate that the Standard Model (SM) is an effective field theory, valid up to a scale $\Lambda$. The excellent agreement between the theory and experiment fixes this scale on a few TeV. Above this scale a new theory must solve the hierarchy problem. One possible solution to this problem was recently given by the called Little Higgs models \cite{AHN}. In these models, the Standard Model Higgs particle is viewed as a pseudo-Goldstone boson of a new global symmetry group. In a first stage, this symmetry is spontaneously broken and the Higgs is a massless particle. A new "collective" symmetry breaking gives a mass to the Higgs boson. The net result is that the new particles and couplings cancel exactly the awful behavior of the Standard Model divergent diagrams giving a light mass to the Higgs fields. Recently, several proposals for "Little Higgs" models have been made. Their main difference are the choice of the new global symmetry. One explicit model \cite{AHQ}, with fewer number of new parameters as possible, was named as the "Littlest Higgs model" (LHM). It is constructed using an $ SU(5)/S0(5)$ coset: a gauged $(SU(2)\otimes U(1))^2$ is broken to its subgroup $SU(2)_L \otimes U(1)_Y$. A detailed study of this model was done in reference \cite{HWU} and some more general reviews are given in \cite{Sch,Per}. Many proposed models predict the existence of an extra heavy neutral gauge boson,  $Z^\prime$. These include: 3-3-1 models \cite{PIV,FRA}, little Higgs model \cite{LIT}, left-right symmetric models \cite{LRM}, supersymmetric breaking with an extra nonanomalous U(1) \cite{MIR}, superstring inspired E$6$ model \cite{E6M} and models with extra dimensions as Kaluza-Klein excitations of neutral gauge bosons \cite{RIZ}. 
\par
The experimental verification of these models, and alternatives like supersymmetric models, could be done by the small quantum corrections to the SM predictions and/or by the discovery of new particles. This will be the main task of the high energy and luminosity accelerators LHC and ILC. Since there are many models with new interactions and particles, it will be necessary to have very clear signatures to distinguish the available models. From the experimental point of view a fundamental question is the nature and properties of the lightest particle in extended models. In the LHM this  particle is a new neutral gauge boson, named $A_H$. A simple and unique property of this model is the presence of a second, heavier, new neutral gauge boson, named $Z_H$ with correlated masses. These new neutral heavy gauge bosons ( named in general here as $Z^{\prime}$ ) could be found in the next generation of high energy colliders: hadron $+$  hadron $ \longrightarrow Z^{\prime} \longrightarrow \ell^+ + \ell^- $and $ e^+ + e^- \longrightarrow Z^{\prime} \longrightarrow \ell^+ + \ell^- $ ( with $ \ell =e, \, \mu \, ,\tau$ ). These topics have been studied by many authors during the last years \cite{NLC,TES}. It is expected that at the LHC energies the new neutral gauge bosons will give detectable signals or new mass limits for the $Z^{\prime}$ and that at future high energy electron-positron colliders it will be possible to study its properties, in details.
\par
The center-of-mass energy of the future colliders will be set to the highest possible values,
consistent with technical feasibility and cost limits.
However it is also possible that the mass of the lightest "Little Higgs" particle will be below the initially designed collider energy. As cross sections at the resonant energies have the highest possible values it will be necessary to redesign the whole collider in order to fix its energy at the new resonant mass. Although physically well motivated, this procedure could be time-consuming and costly. It is important to have alternative signatures and show model differences already in the initial collider energy. Earlier works on the LHM have started from the possibility of a first experimental signature for new gauge interactions at the LHC and then the leptonic colliders were considered at a center of mass energy equal to the new boson mass values \cite{LUA}. We will consider in this paper the possible scenario of having the first new indications of a new neutral gauge boson in lepton colliders. Due to the energy scan difficulties for these kinds of accelerators, we will work with some new neutral boson masses values above and below at a fixed lepton collider energy.
\par
The main point proposed in this paper is that it is not necessary to wait until the collider energy could be fixed at the mass of the new neutral gauge boson in order to get its properties and exclude some models. We will consider new bounds and model differences of new neutral gauge bosons for different $\sqrt s$ values, taking $\sqrt s \not= M_Z^{\prime}$, both for polarized and unpolarized beams. 
\par
In this paper we will discuss two main signatures for electron-positron colliders that can point to model differences: asymmetries in the production of fermion pairs and the production of a single hard photon associated with new heavy neutral gauge bosons. Cross sections will be shown to present detectable effects. The fermion pair production allows one to find new limits that can be model dependent for $M_{Z^ \prime} > \sqrt{s}$ and the hard photon production allows one to study the region for $M_{Z^ \prime} < \sqrt{s}$
\par
We concentrate our attention first on the lightest neutral gauge boson of the LHM - $ A_H$. The present experimental limit is $M_{A_H} > 600$ GeV obtained by \cite{PDG} and is somewhat model dependent.  The other neutral gauge boson $ Z_H$ is heavier, with a mass of the order $ M_{Z_H} \simeq 4 M_{A_H}$.
\par
In order to compare the LHM predictions with other models we will employ the canonical $ \eta ,\ \chi , \psi $ superstring inspired $E_6$ models \cite{PDG}, and Symmetric and Mirror left-right models \cite{NOS}. As our main interest is the detection of new gauge bosons, we will consider only their coupling to ordinary matter. New particle states must also be coupled to new gauge bosons but this will introduce new parameters and will make the estimates more model dependent.
\par
The general form of the interaction Lagrangian involving only ordinary fermions and extra neutral gauge bosons is given by
\begin{equation}
{\cal L}= \bar\Psi\gamma_{\mu}(g_V - g_A\gamma_5)\Psi Z^{\prime\mu}.
\end{equation}
\noindent
where $g_V$ and $g_A$ are respectively the vector and vector-axial couplings. The $\gamma$ and $Z$ couplings to the ordinary matter are the same as the SM ones.
\par
For the LHM, according to reference \cite{HWU}, the new neutral gauge bosons
$A_H$ and $Z_H$ couple with ordinary matter with mixing angles $\theta^{\prime}$ and $\theta$ respectively.

The $\eta, \chi ,\psi $ superstring inspired $E_6$ models \cite{PDG} and the Symmetrical and Mirror left-right models [17]  have a new neutral gauge boson $Z^{\prime}$ only coupled to fermions.

\par
The angular asymmetry is an interesting variable because it is strongly dependent on the couplings $g_V$ and $g_A$.
The leptonic asymmetries for the process $e^+ + e^- \longrightarrow  \mu^+ + \mu^- $ presented in \cite{CHI,CHO} were calculated for $\sqrt s\simeq M_{A_H}$ where $M_{A_H}$ is the lightest new neutral boson mass. In this paper, we consider the same channel but for new gauge boson mass regions below and above the collider energy.
In all calculations, each final lepton was required to be detected within the azimuthal angle range $\vert\cos{\phi}\vert < 0.995$, where $\phi$ is the angle of either of the final fermions with respect to the direction of the electron beam. This cut corresponds roughly to the detector limitations.
For the LHM, the total width $\Gamma_{A_H}$ was calculated considering the main contributions from pairs of ordinary charged and neutral fermions  and
$A_{H}$-$Z$-$H$ couplings.  For the $\Gamma_{Z_H}$ width we have included these channels and the relevant contribution from $W^+ W^-$ channel, which is negligible for $A_H$. The mass relation between
$A_H$ and $Z_H$ is approximately given by the relation $M_{Z_H}\simeq\sqrt{5}M_{A_H}/\tan\theta_W$. As a consequence, in the energy region from $500$ GeV to $1$ TeV the processes will be dominated by $Z$ and $A_H$ exchange, if we assume $M_{A_H}> 500$ GeV. Typical values for the widths, for a choice  $c\equiv\cos\theta=0.3$, $c'\equiv\cos\theta'=0.62$, give us
$\Gamma_{Z_H}=11 $ GeV for $M_{Z_H}= 2.4$ TeV and $\Gamma_{A_H}=0.07$ GeV for $M_{A_H}=500$ GeV, in accordance with the results in reference \cite{HWU}.
A first result involving LHM parameters is the total cross section for $e^+ + e^- \longrightarrow\mu^+ + \mu^- $ as a function of the $\sqrt s$, for $M_{A_H}= 500$ GeV and $M_{A_H}= 800$ GeV. The curves are shown in Figure 1. We have considered the mixing angles $\theta$ and $\theta'$ in the same range as estimated in reference \cite{HWU}. As expected, near the resonant region there is a peak related to the extra gauge boson exchange.
\begin{figurehere}
\begin{center}
\includegraphics[width=5cm,height=4cm]{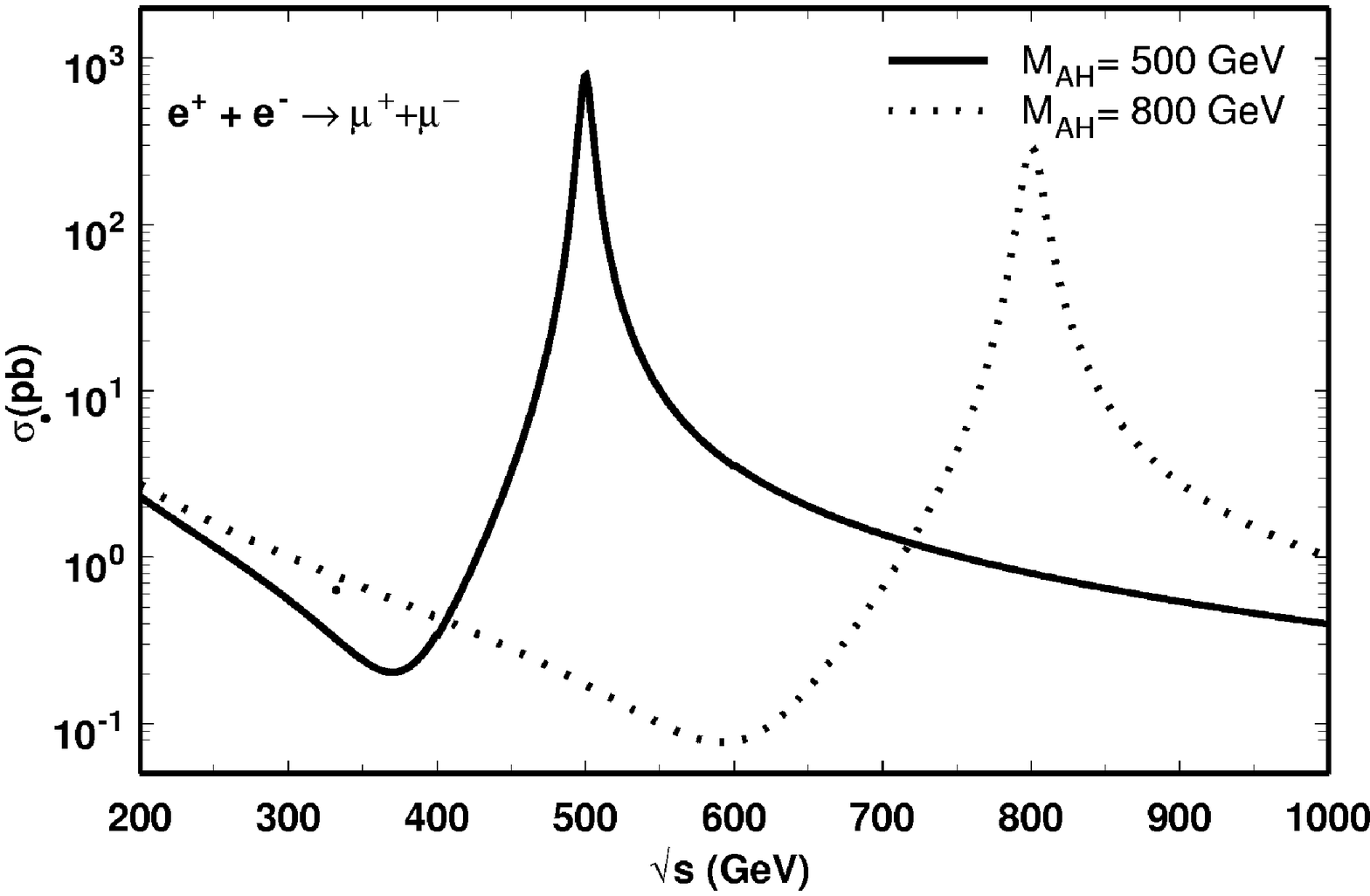}
\caption{Total cross section versus $\sqrt s$ for $M_{A_H}= 500$ and $800$ GeV for LHM ($c = 0.3$ and $c' = 0.71$).}
\end{center}
\end{figurehere}
A first difference between the theoretical models can be found in the forward-backward asymmetries ($A_{FB}$) in $ e^+ + e^- \longrightarrow f + \bar f $, with $f=\mu^-, c$ and $b$. These fermion couplings to $A_H$ allow a clear model separation from other extended models. We have performed the calculation with the CompHep package \cite{HEP}, in which the previously mentioned models were implemented. The curves for $A_{FB}$ are displayed in Figures 2, 3 and 4 for the models cited above: LHM with a choice of the mixing parameters $c$, $c'$; superstring inspired models $\eta ,\chi ,\psi $ and Symmetric and Mirror left-right models as a function of $M_{Z^{\prime}}$ (and $M_{A_H}$) for $\sqrt s= 500$ GeV. Near the resonant region ( $\sqrt s= 500$ GeV ) some models are more sensitive than the LHM, presenting sizeable deviations from the SM value $A_{FB}^{SM}=0.47$. The same behavior occurs for $\sqrt s= 1$ TeV. For larger values of the mass of the new gauge boson, the combined analysis of $\mu$, $c$ and $b$ pair production might help to stablish the best model underlying these deviations from the SM predictions.
\begin{figurehere}
\begin{center}
\includegraphics[width=5cm,height=4cm]{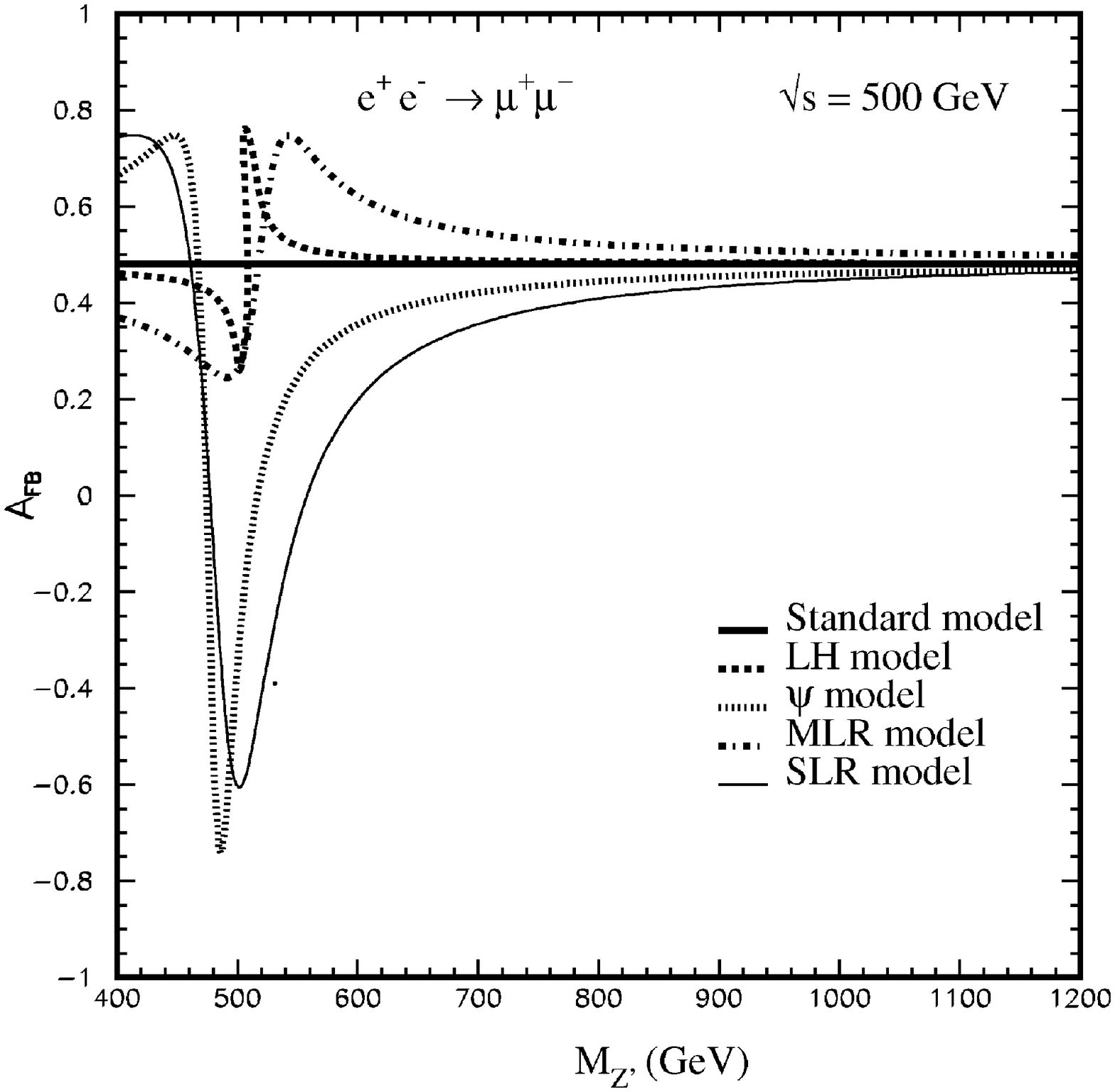}
\caption{Forward-backward asymmetry versus $M_{Z^{\prime}}$ ($M_{A_H}$) for $\sqrt s=500$ GeV in $e^+ + e^- \lra \mu^+ +\mu^-$ for some models.}
\end{center}
\end{figurehere}
\begin{figurehere}
\begin{center}
\includegraphics[width=5cm,height=4cm]{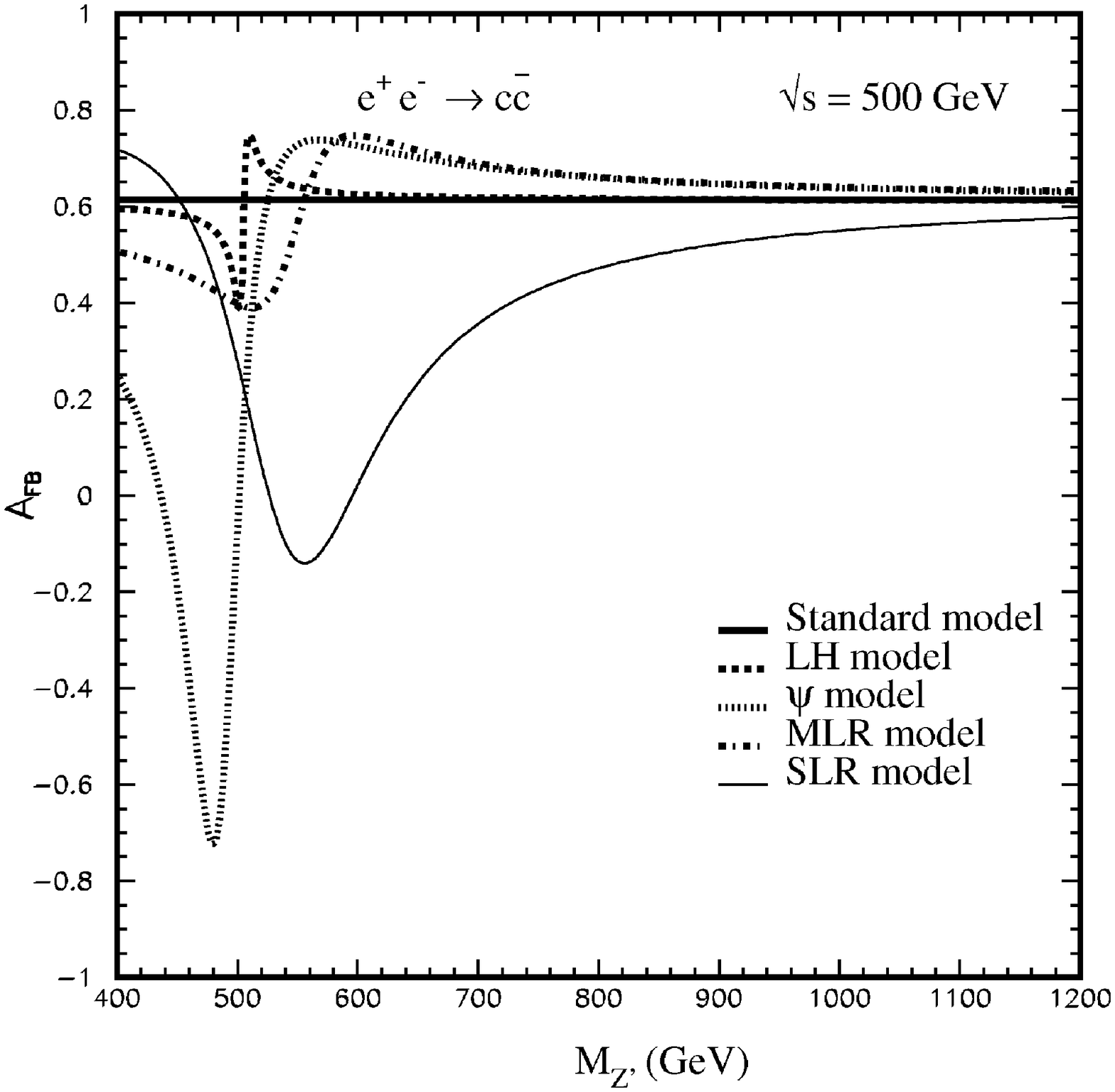}
\caption{Forward-backward asymmetry versus $M_{Z^{\prime}}$ ($M_{A_H}$) for $\sqrt s=500$ GeV in $e^+ + e^- \lra c +\bar c$ for some models.}
\end{center}
\end{figurehere}
\begin{figurehere}
\begin{center}
\includegraphics[width=5cm,height=4cm]{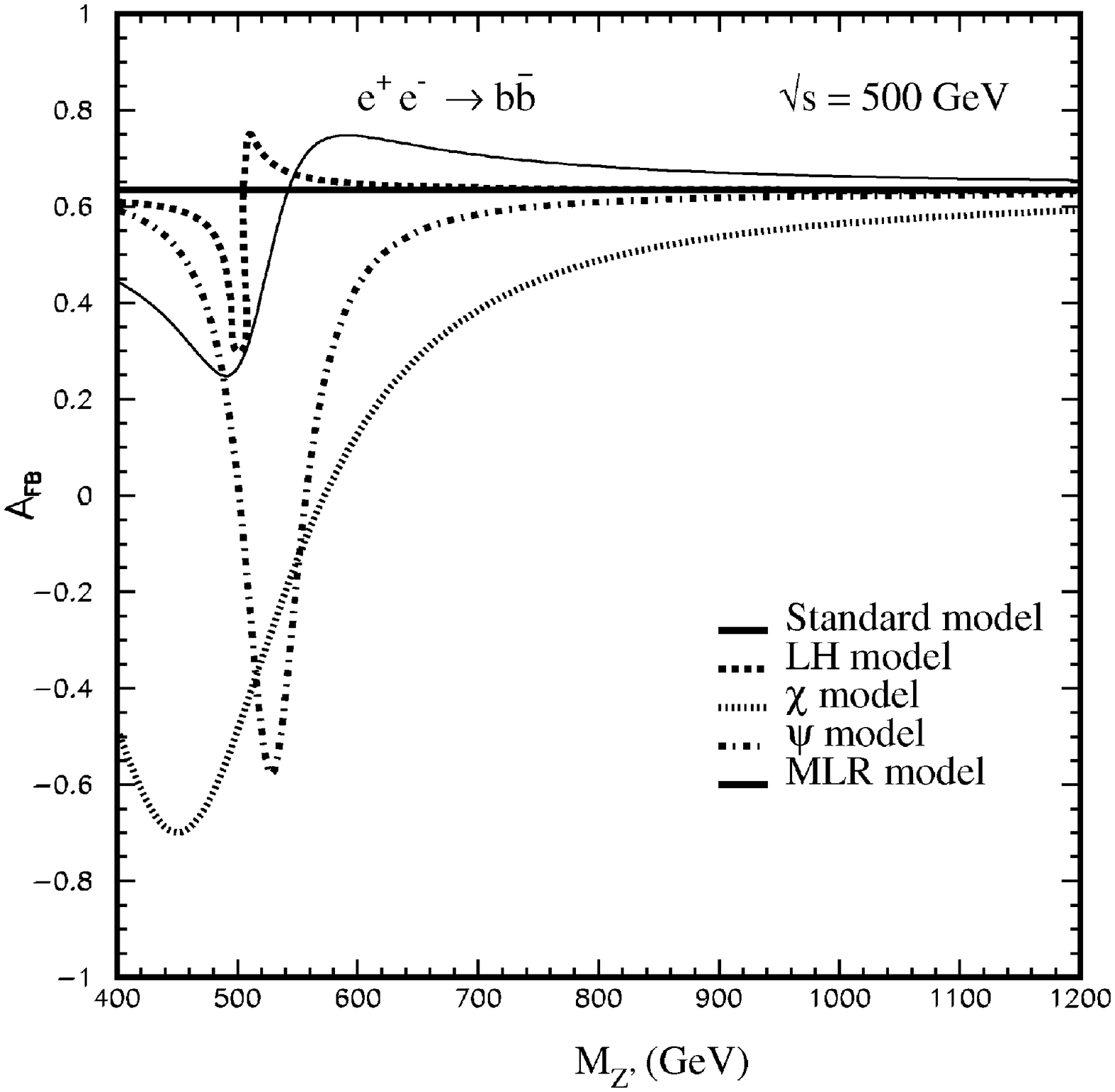}
\caption{Forward-backward asymmetry versus $M_{Z^{\prime}}$ ($M_{A_H}$) for $\sqrt s=500$ GeV in $e^+ + e^- \lra b +\bar b$ for some models.}
\end{center}
\end{figurehere}
Another variable that shows differences between models is the left-right asymmetry ($A_{LR}$). 
We present in Figure 5 for $\mu^+ \mu^-$, Figure 6 for $c \bar c$ and Figure 7 for $b \bar b$ the left-right asymmetry for different models as a function of $M_{Z^{\prime}}$ for $\sqrt s= 500$ GeV. Again the $A_{LR}$ for the LHM has a small deviation from the SM value. The $A_{LR}$ for all models has a smaller deviation from the SM than the $A_{FB}$ for the corresponding models.

\begin{figurehere}
\begin{center}
\includegraphics[width=5cm,height=4cm]{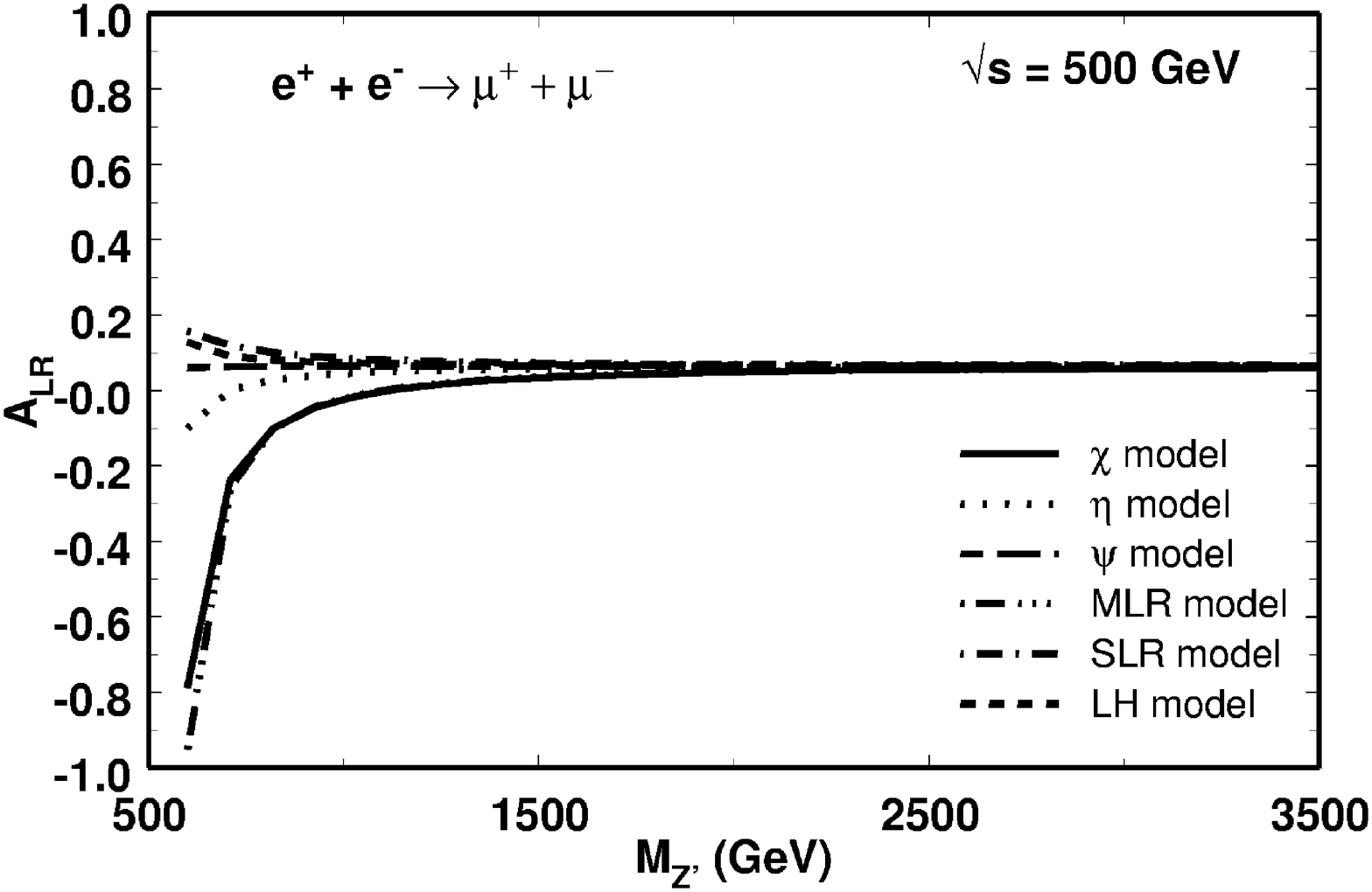}
\caption{Left-right asymmetry in $e^+ + e^- \lra \mu^+ +\mu^-$ versus $M_{Z^{\prime}}$ ($M_{A_H}$) for $\sqrt s= 500$ GeV for some models.}
\end{center}
\end{figurehere}

\begin{figurehere}
\begin{center}
\includegraphics[width=5cm,height=4cm]{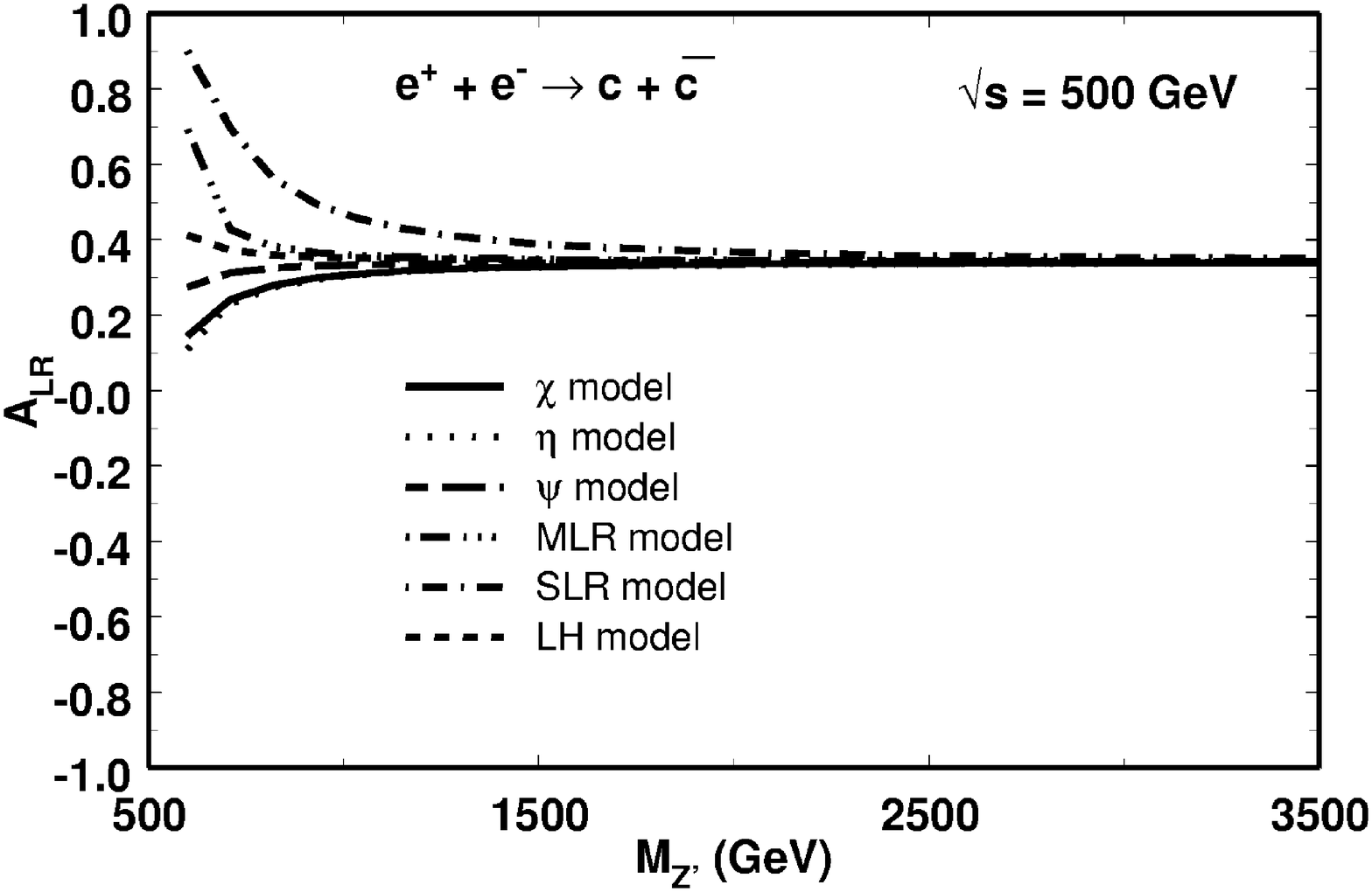}
\caption{Left-right asymmetry in $e^+ + e^- \lra c +\bar c$ versus $M_{Z^{\prime}}$ ($M_{A_H}$) for $\sqrt s= 500$ GeV for some models.}
\end{center}
\end{figurehere}

\begin{figurehere}
\begin{center}
\includegraphics[width=5cm,height=4cm]{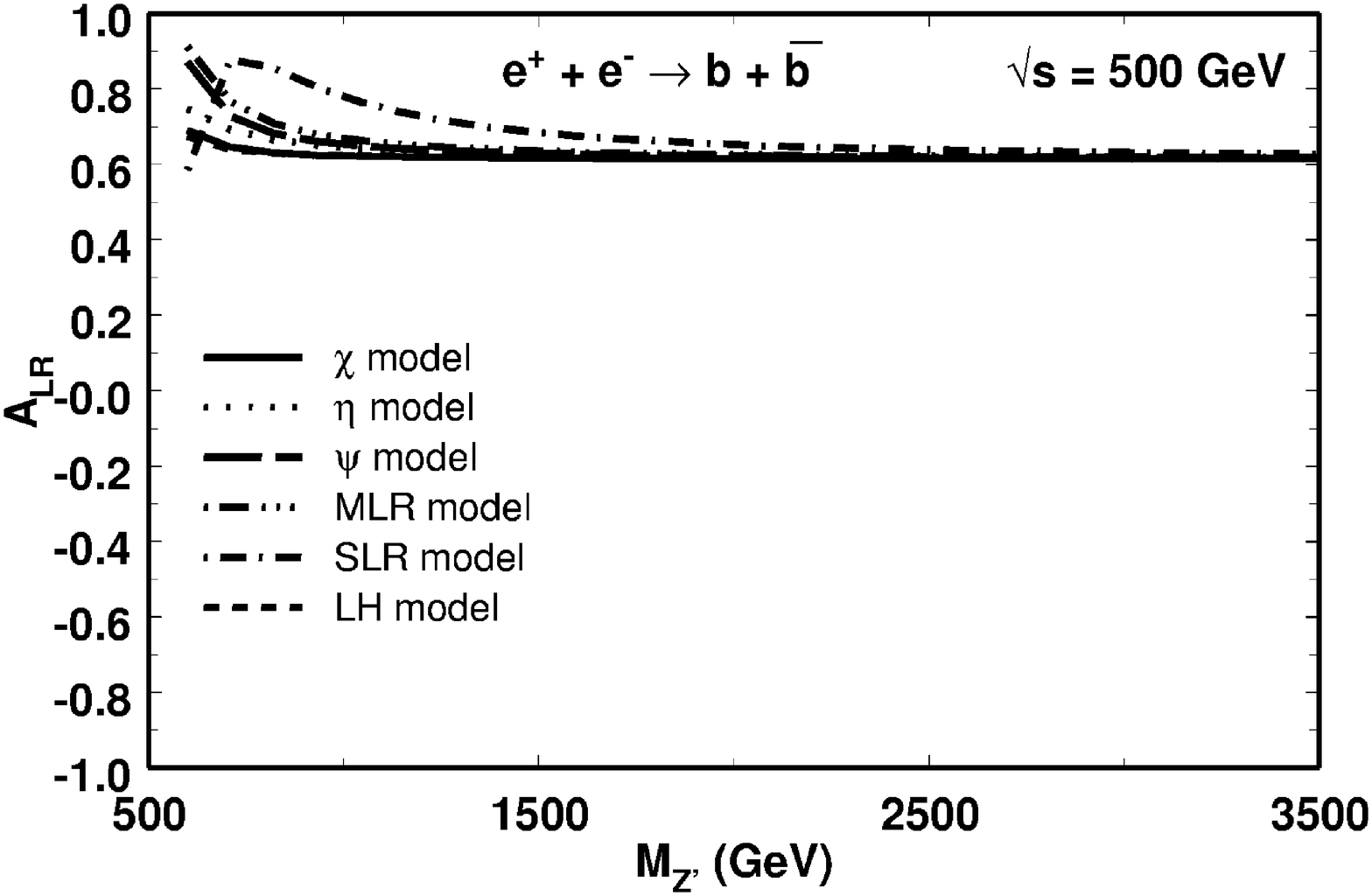}
\caption{Left-right asymmetry in $e^+ + e^- \lra b +\bar b$ versus $M_{Z^{\prime}}$ ($M_{A_H}$) for $\sqrt s= 500$ GeV for some models.}
\end{center}
\end{figurehere}
Bounds on the LHM were previously obtained in \cite{HWU,CHI,CHO,GOD} at the energy peak of the new neutral gauge boson production, using a combination of physical observables at that energy. Deviations from the SM predictions were shown to be within these bounds.

In this paper we present new bounds on the mass $M_{A_H}$ of the new neutral gauge boson $A_H$ 
as a function of $\cos \theta^{\prime}$. These bounds are derived from the angular distribution of the final-state fermion, by means of a $\chi^2$ test. Assuming that the experimental data in the fermion-pair production will be described by the SM predictions, we defined a one-parameter $\chi^2$ estimator
\begin{equation}
{ \chi^2 = \sum_{i=1}^{n_b} {\biggl( {N_i^{SM}- N_i^{LHM} \over
\Delta N_i^{SM}}\biggr)^2}},
\end{equation}
\noindent where $N_i^{SM}$ is the number of SM events collected in the
$i^{th}$ bin, $N_i^{LHM}$ is the number of events in the $i^{th}$ bin as
predicted by the LHM, and $\Delta N_i^{SM} =
\sqrt{(\sqrt {N_i^{SM}})^2 + (N_i^{SM}\epsilon)^2}$ the corresponding
total error, which combines in quadrature the Poisson-distributed statistical
error with the systematic error. We took $\epsilon = 5\%$ as the systematic error
in our calculation. We considered the muon, charm and bottom detection efficiencies equal to $95\%$, $60\%$ and $35\%$ respectively. We estimated upper bounds for $M_{A_H}$ with $95\%$ C.L., fixing $c = 0.3$ and using the mixing angle $\theta^{\prime}$ as a free parameter. Figure 8 shows the bounds on $M_{A_H}$, for $\sqrt s= 500$ GeV and an integrated luminosity ${\cal L}= 100$ fb$^{-1}$. The $95\%$ C.L. bounds for $\sqrt s= 1$ TeV and ${\cal L}= 340$ fb$^{-1}$ are displayed in Figure 9. In both cases the muon channel was found to be more restrictive than the hadron channels. 
\begin{figurehere}
\begin{center}
\includegraphics[width=5cm,height=4cm]{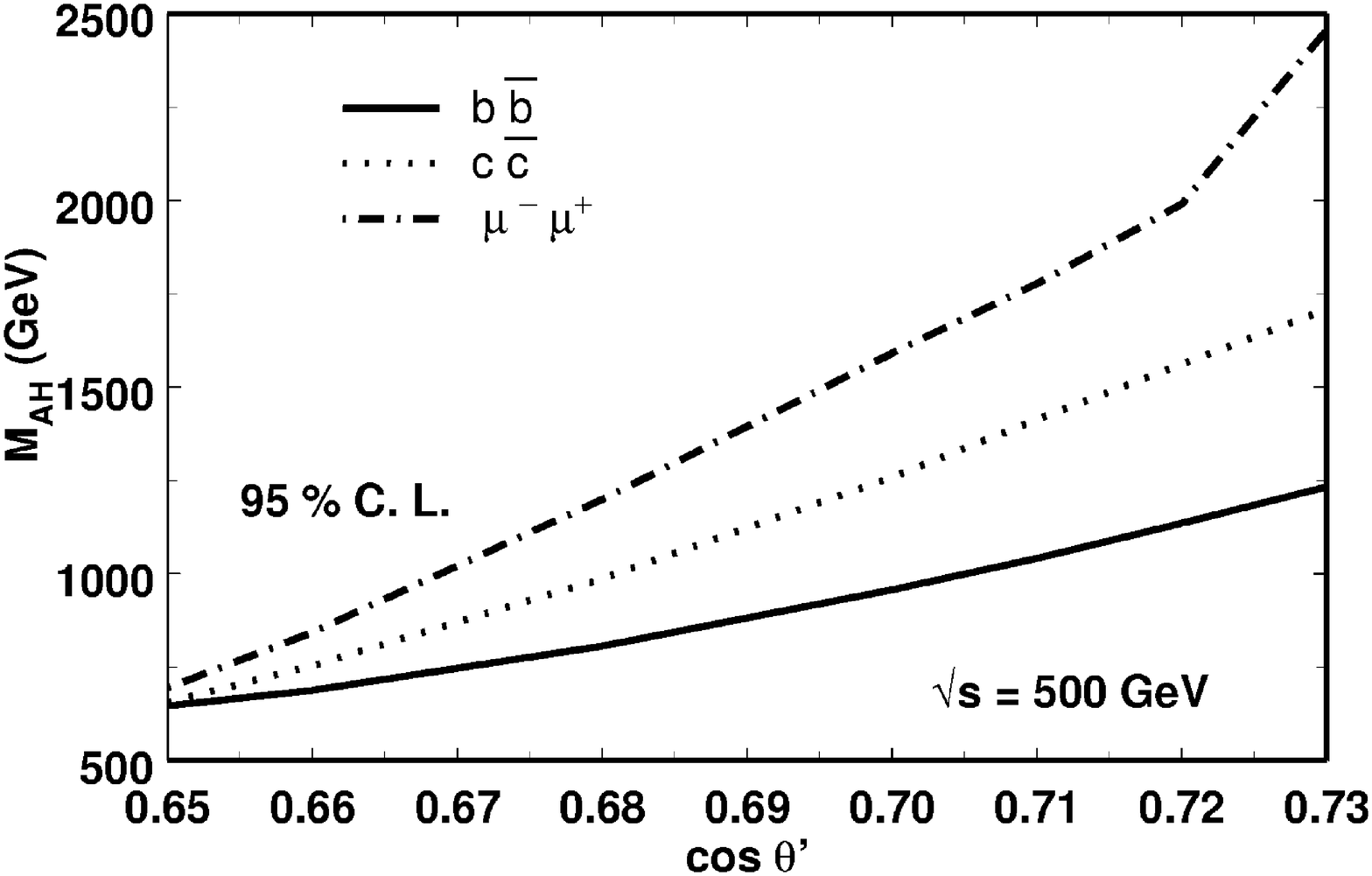}
\caption{ $M_{A_H}$ bounds ($95\%$ C.L.) as a function of $\cos \theta'$ in $e^+ + e^- \lra f +\bar f$, where $f=\mu,  c$ and $b$ for $\sqrt s= 500$ GeV for LHM.}
\end{center}
\end{figurehere}
\begin{figurehere}
\begin{center}
\includegraphics[width=5cm,height=4cm]{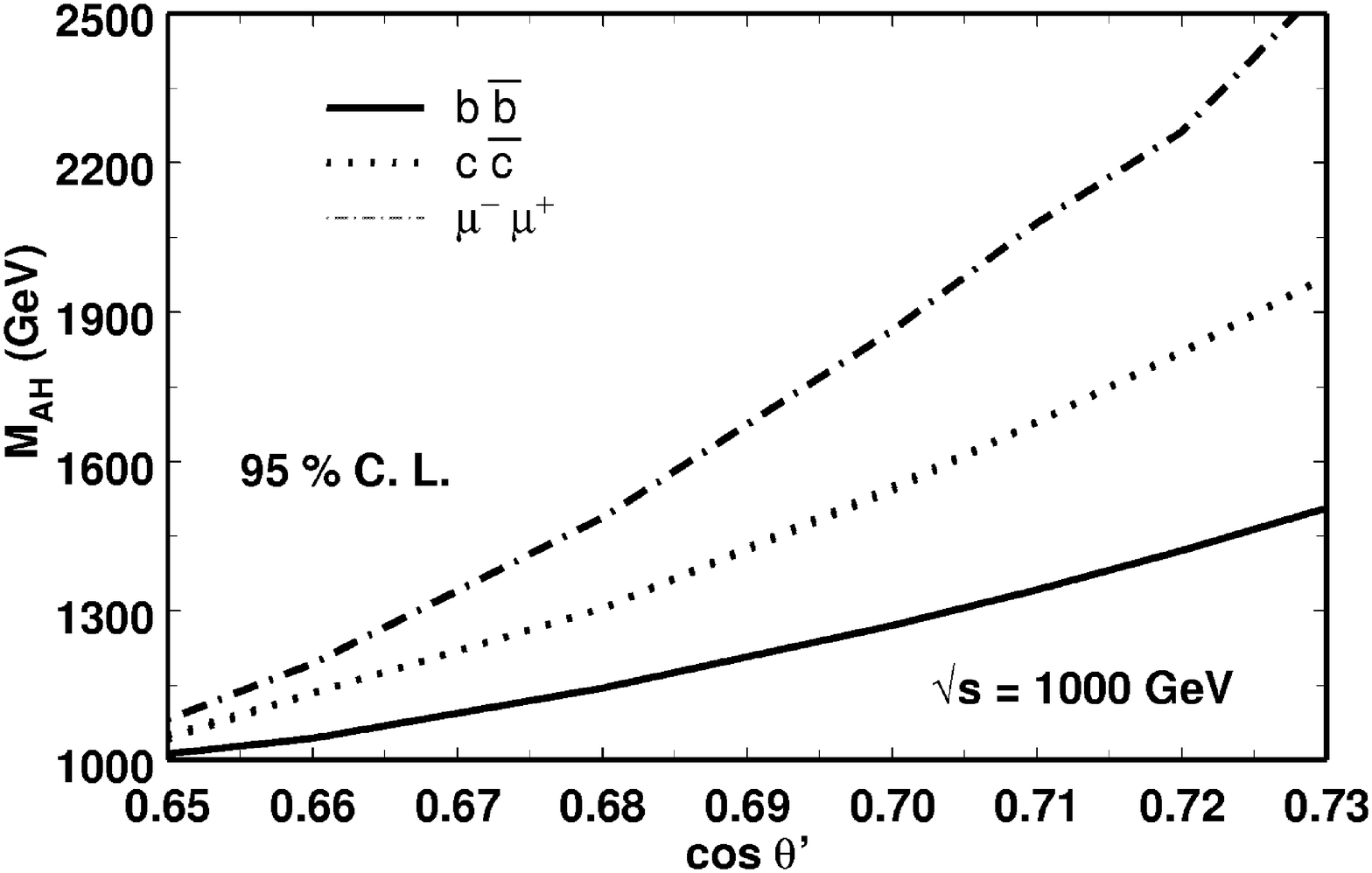}
\caption{$M_{A_H}$  bounds ($95\%$ C.L.) as a function of $\cos \theta'$ in $e^+ + e^- \lra f +\bar f$, where $f=\mu,  c$ and $b$ for $\sqrt s= 1$ TeV for LHM.}
\end{center}
\end{figurehere}
\begin{figurehere}
\begin{center}
\includegraphics[width=5cm,height=4cm]{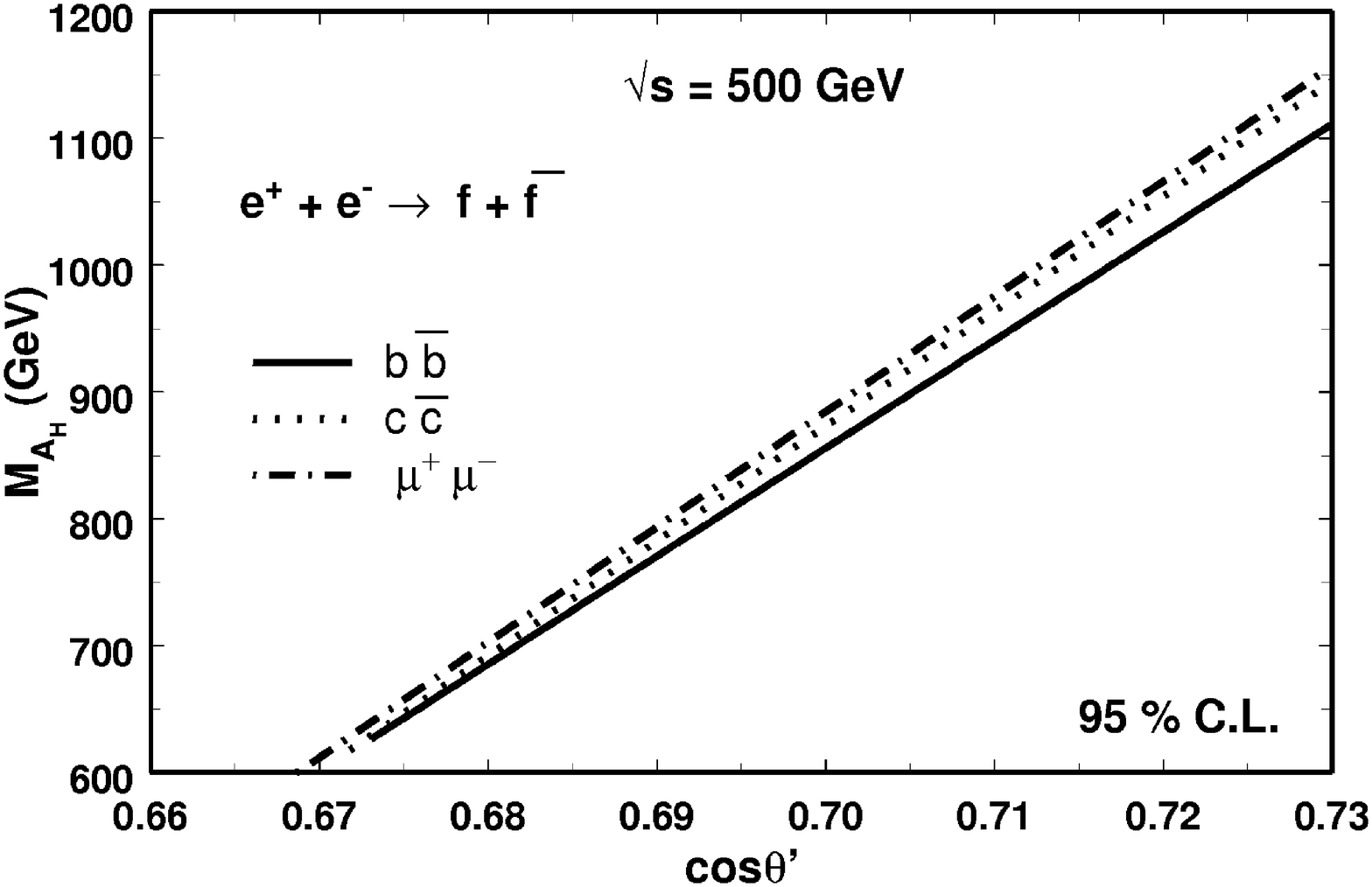}
\caption{$M_{A_H}$ bounds ($95\%$ C.L.) as a function of $\cos\theta'$ in $e^+ + e^- \lra f +\bar f$, where $f=\mu, c$ and $b$ for $\sqrt s= 500$ GeV for LHM, considering polarized beams.}
\end{center}
\end{figurehere}
In Figure 10 we display the resulting limits on $M_{A_H}$ for $\sqrt s= 500$ GeV, but considering longitudinally polarized beams. The degrees of polarization of the electron and
positron beams were taken to be $90\%$ and $60\%$ respectively. The limits obtained in this case
are less restrictive than the corresponding bounds derived from unpolarized angular distributions, and are essentially the same for all three final states.
\par
A second point that can show interesting properties of a new fundamental interaction mechanism is the associated production of a general $Z^{\prime}$ and a hard photon \cite{NOS} in the process $e^+ + e^- \longrightarrow \mu^+ + \mu^- + \gamma $. This channel is suitable when $M_{Z^\prime} < \sqrt s$. The main advantage of this channel is to show that one can study the new $Z^\prime$ properties before all the redefinition of the collider energy at the new neutral gauge boson mass. The cross section for this channel is smaller than the direct resonant cross section but can give a statistically meaningful number of events, since the luminosity will be larger than $100$ fb$^{-1}$. Another important advantage is that one does not need to reconstruct the heavy neutral boson mass from its decay products. The direct $Z^\prime$ decay is important, but the reconstruction efficiency will reduce the total number of events. This reconstruction can be more complicated when these new bosons decay into unstable particles, like hadrons.
\par
 The hard photon emission proposed in this paper is very different from the usual multiphoton emission accompanying $Z^\prime$ production. The well known logarithm corrections \cite {NIC} were studied more recently \cite{FRE} and imply changes at the $Z^\prime$ pole. But these corrections are different from the kinematics of two body final state in the associated hard photon production \cite{NOS}.
\par
A very simple consequence of the conservation of energy and momentum, is that the final high energy hard photon has a fixed energy given by
\begin{equation}
E_{\gamma}\mp \Delta_{\gamma}=\frac{s - ({M_{Z^{\prime}}}\pm \Delta_{Z^{\prime}})^2}{2 \sqrt s}
\end{equation}
where $\Delta_{\gamma}$ and $\Delta_{Z^{\prime}}$ are the uncertainties in the photon energy and $M_{Z^{\prime}}$.
\par
The study of the hard photon energy distribution will give the same information as the direct $Z^{\prime}$ decays, but in a simple and direct way. The resulting photon energy distribution contains important information about the $A_H$ and $Z_H$ bosons and is shown in Figures 11 and 12. For $\sqrt s=3$ TeV if $M_{A_H}< 750$ GeV, the distribution $d\sigma/dE_{\gamma}$ shows two peaks, coming from $A_H$ and $Z_H$, when considering the LHM. For $M_{A_H} > 750$ GeV only the peak associated to $A_H$ will show up. The figures were obtained using a cut $E_{\gamma}> 50$ GeV on the hard photon energy, $E_j > 5$ GeV and $\vert \cos \phi_i\vert < 0.995$, where $j = f, \bar f$ and i = $\gamma, f \bar f$. The $E_{\gamma}> 50$ GeV cut eliminates the contributions from the $\gamma$'s soft emissions.

We notice the strong mixing-angle dependence on the magnitude of the photon peaks in Figure 11. The peak associated to $Z_H$ will depend on $\cos \theta$ and the peak associated to $A_H$ will vary with $\cos\theta^\prime$.  The $\eta,\chi$ and $\psi $ models have a much larger value for the total neutral gauge boson width and this fact makes the hard photon energy much broader as shown in Figure 12. The $E_{\gamma}$ distribution allows us to find the $Z^{\prime}$ energy and to distinguish the models, in connection with Eq. (3).

\begin{figurehere}
\begin{center}
\includegraphics[width=5cm,height=4cm]{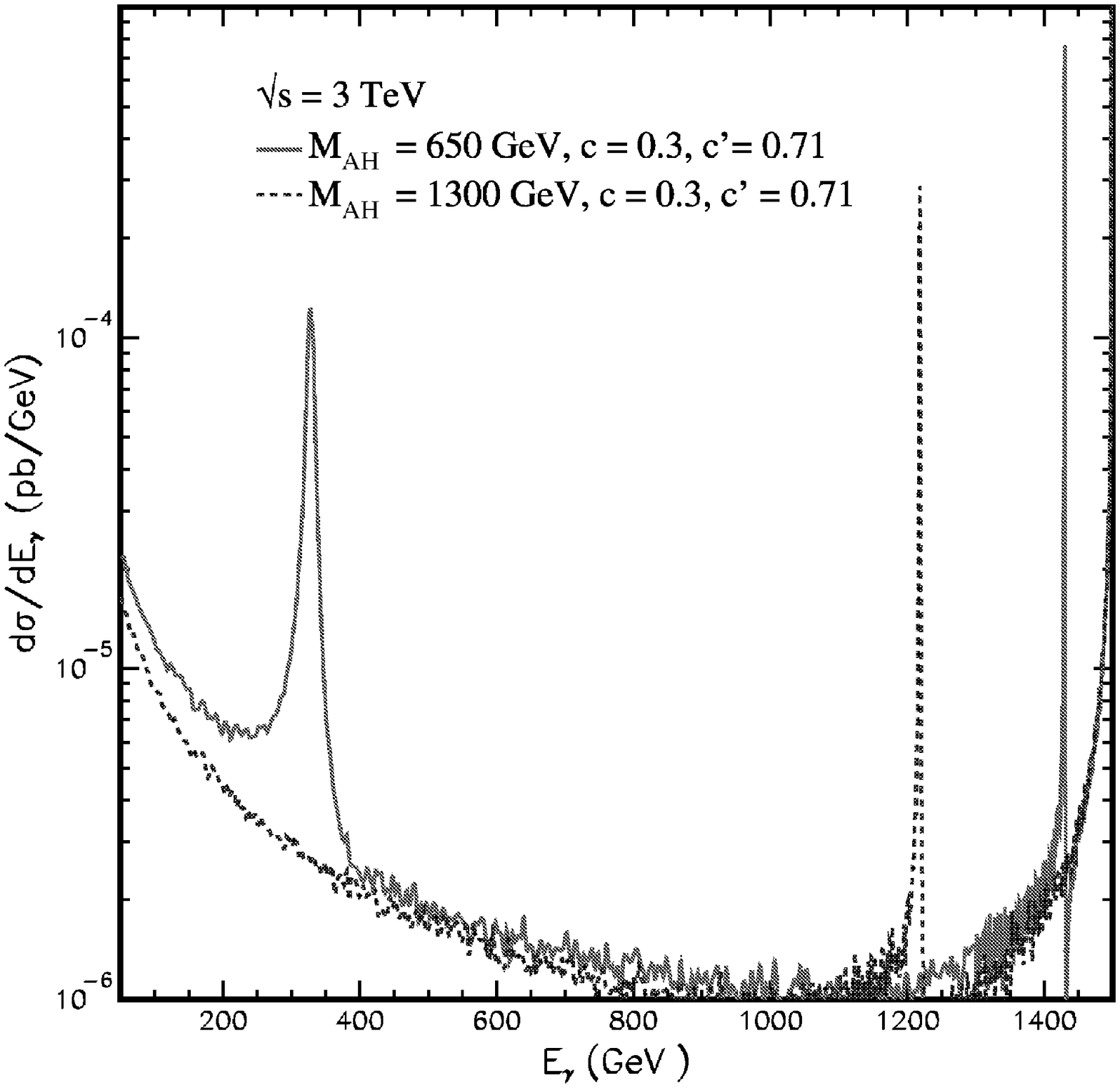}
\caption{Photon energy distribution in $e^+ + e^- \lra \gamma +\mu^- +\mu^+$ showing  two peaks associated to $Z_H$ and $A_H$ with $\sqrt{s} = 3$ TeV for LHM ($c = 0.3$ and $c' = 0.71$), when $M_{A_H}= 650$ GeV and only one peak when $M_{A_H}= 1300$ GeV .}
\end{center}
\end{figurehere}

\begin{figurehere}
\begin{center}
\includegraphics[width=5cm,height=4cm]{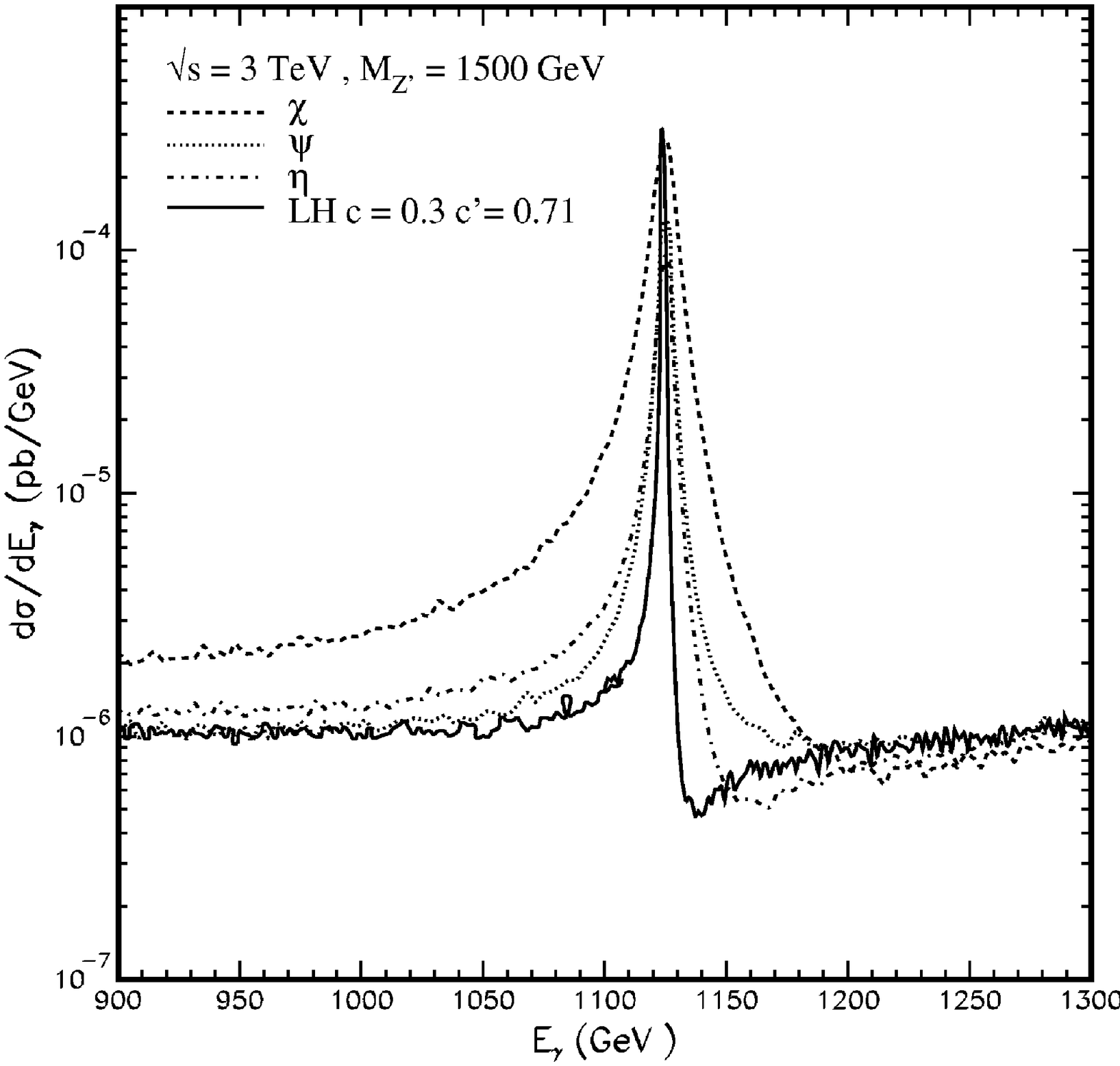}
\caption{Photon energy distribution in $e^+ + e^- \lra \gamma + \mu^- +\mu^+$ with $\sqrt{s} = 3$ TeV for $\chi$,  $\eta$,  $\psi$ and for LHM ($c = 0.3$ and $c' = 0.71$).}
\end{center}
\end{figurehere}

\par
In conclusion, we have shown that $A_{FB}$ and $A_{LR}$ in $ e^+ + e^- \longrightarrow  f^+ + f^- $, where $f = \mu$, $b$ and $c$, for $M_{Z^\prime} > \sqrt{s}$, can distinguish the LHM from other proposed models in a wide mass range.
A  $\chi^2$ analysis on the angular distribution of the final fermions for the same channels allows us to obtain the mass limits as a function of $\cos\theta^\prime$. The above calculations were done considering polarized and unpolarized beams. Using the angular distribution it is possible to get other mass limits than using the total cross-sections as done by \cite{CHI}.
\par
For $M_{Z^\prime} < \sqrt{s}$, the hard photon energy distribution in $ e^+ + e^- \longrightarrow\gamma + \mu^+ + \mu^- $ can indicate the theoretical origin of new possible heavy neutral gauge bosons independent of
how these new bosons decay. This alternative signature for $Z^\prime$  production at the
new electron-positron colliders could allow us to study its properties, at a
fixed collider energy.
If $ M_{A_H} < \sqrt{s} < M_{Z_H}$  only one peak shows up in the hard photon energy distribution, and the models can be distinguished using the width of this peak.
If $M_{Z_H}<\sqrt{s}$ two peaks appear in the photon energy distribution, and all the models will be excluded, except the LHM.

In order to detect a new gauge boson and understand its properties, the center-of-mass energy of the future high-energy electron-positron experiments will have to be set to the mass of this new boson. The main point presented in our work is to show that we can establish some properties of the possible new gauge boson already at the initial collider energy.
\par
\noindent {\it Acknowledgments:} This work was partially supported by the
following Brazilian agencies: CNPq, FAPEMIG and FAPERJ.

\end{multicols}
\end{document}